# Cross-phase modulation instability in mode-locked laser based on reduced graphene oxide

Lei Gao, *Student Member, IEEE*, Tao Zhu, *Member, IEEE*, Min Liu, and Wei Huang

***Abstract*—Cross-phase modulation instability (XPMI) is experimentally observed in a fiber ring cavity with net normal dispersion and mode-locked by long fiber taper. The taper is deposited with reduced graphene oxide, which can decrease the threshold of XPMI due to the enhanced nonlinearity realized by 8 mm evanescent field interaction length and strong mode confinement. Experimental results indicate that the phase matching conditions in two polarization directions are different, and sidebands with different intensities are generated. This phase matching condition can be satisfied even the polarization state of the laser varies greatly under different pump strengths.**

## I. INTRODUCTION

MODULATION instability (MI) in nonlinear and dispersive media is one of the fundamental topics in plasma physics [1], fluid dynamics [2], Bose-Einstein condensates and nonlinear optics [3-5], where the steady state of a system is destabilized by weak perturbations [6]. MI in optical fiber has drawn considerable attentions for potential applications in ultra-short lasers with high repetition rate, optical parametric amplifiers and all optical logic devices [4-9], where sidebands grow exponentially as the energy exchanges between the pumping wavelength and MI gain [7]. Physically, it can be interpreted as degenerate four wave mixing (DFWM), in which two pump photons ($\omega_P$) with the same frequency are annihilated and two symmetrical new photons at Stokes frequency ($\omega_S$) and anti-Stokes frequency ($\omega_{AS}$) are created simultaneously. Both the energy conservation and phase matching are satisfied [6].

The cross-phase modulation MI (XPMI) involves the interaction of two orthogonal polarizations, and the sidebands are approximately polarized 45° to that of pump laser [3,10]. It can be observed in fibers with anomalous and normal group velocity dispersion (GVD), and the phase matching process is shown in Fig. 1.

However, due to the relative weak nonlinearity, MI in conventional fiber often requires high pump peak power [3,6,7].

This work was supported by Natural Science Foundation of China (No. 61377066 and 61405020), the Fundamental Research Funds for the Central Universities (No. 106112013CDJZR120002 and 106112013CDJZR160006).

Lei Gao, Tao Zhu, Min Liu, Wei Huang are with the Key Laboratory of Optoelectronic Technology & Systems (Education Ministry of China), Chongqing University, Chongqing 400044, China. (Corresponding email: zhutao@cqu.edu.cn)

To decrease the MI threshold, researchers have focused on MI in fibers with periodical variations of dispersion or nonlinearity [11-14], and MI in fiber resonators, where parametric gain is increased due to accumulated nonlinearity experienced by pulses circulating in laser cavity [15-17]. Haelterman *et al* show that MI could occur in a fiber ring resonator with normal dispersion, where the phase matching was obtained by the coupling between two polarization components [15]. Tang *et al* reported new spectral components in passively mode-locked fiber laser [16], where high pump power leads to unstable dispersive waves [16]. Luo *et al* observed MI with subsidebands induced by periodic power variation [17].

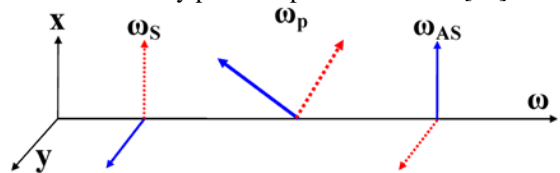

Fig. 1 Phase matching process of XPMI. The solid lines and dotted lines represent two orthogonal polarization components.

In order to decrease the threshold of MI, we enhance the nonlinearity in a fiber laser cavity via long evanescent-field interaction length and strong mode confinement. To realize this, we use a long fiber taper deposited with reduced graphene oxide (reGO) [18-22]. The spontaneous XPMI is observed when pump power is set at 185 mW.

## II. EXPERIMENTAL SETUP AND RESULTS

The fiber ring cavity is schematically shown in Fig. 2 (a), where a 9.7 m erbium-doped fiber (EDF, EDF-980-T2) with dispersion parameter of -12.7 ps/nm/km is pumped by 980 nm laser via a wavelength division multiplexer (WDM). Unidirectional operation is forced by a polarization independent isolator, and birefringence is tuned by a mechanical polarization controller (PC). The output is taken by a 10% fiber coupler. The ring cavity includes a 17 m standard single mode fiber (SMF, Corning SMF-28) with dispersion parameter of 18 ps/nm/km and a 19.5 m dispersion compensation fiber (DCF, DCF38) with dispersion parameter of -38 ps/nm/km. The net normal dispersion is a necessary condition to generate MI, while no MI can be found in net anomalous dispersion cavity. All components are polarization-independent.

The reGO is synthesized by reducing graphene oxide, and the D peak and G peak in its Raman spectrum correspond to

~1320 cm$^{-1}$ and ~1610 cm$^{-1}$, respectively. The taper-graphene saturable absorber (SA) is produced via depositing reGO flakes onto the fiber taper via light pressure [23, 24]. The reGO solution has to be dissolved well and centrifuged to get transparent solution, and the injection laser power has to be selected properly for a controllable deposition process. The fiber taper with waist diameter and length of ~7.5 μm and ~8 mm, and the transmission after reGO deposition are shown in Fig. 2 (b). The laser output is recorded by a detector (PDB430C, Thorlabs Co,. Ltd), an oscilloscope (Infiniium MSO 9404A, Agilent Tech.), an optical spectrum analyzer (Q8384, Advantest Corp.) and a polarimeter (PSGA-101-A, General Photonics Corp.).

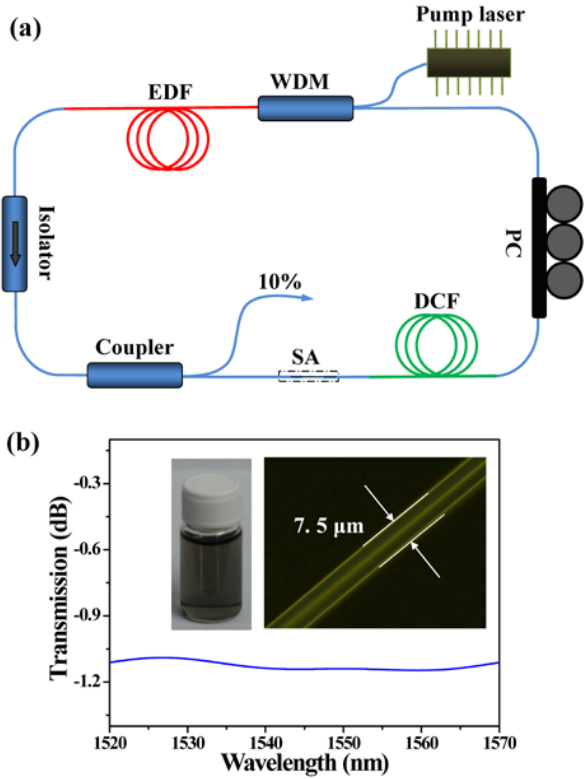

Fig. 2 (a). Schematic of the fiber ring cavity. (b). Transmission of fiber taper deposited by reGO, the insets show the transparent reGO solution and microscope image of the taper.

The mode locking threshold is about 25 mW, and the XPMI occurs when pump strength exceeds 185 mW. The weak sidebands increase exponentially, while the main wavelength almost kept constant. Figure 3(a) shows the typical temporal output for pump power at 80 mW and 320 mW, where pulse durations of the two lasers are 23.6 ns and 4 ns, and the maximum peak power is less than 1 W. The wavelength difference between the sidebands and the main wavelength is about 2.9 nm.

In our experiment, the fiber taper alone could not produce noticeable modulation for MI. This has been verified by replacing the taper-graphene structure with a graphene between two pigtails and a similar fiber taper without reGO deposition, and no noticeable MI can be found. We can conclude that it is the long fiber taper deposited with reGO that attributes to the formation of MI. The nonlinearity is enhanced by the reGO flakes deposited on the 8 mm taper waist. The threshold is decreased by the periodic dispersion and loss variations experienced by pulses circulating in the cavity [15-17]. The corresponding optical spectrum for ML under XPMI shown in Fig. 3(b) exhibits fluctuations in two sidebands, which may originate from the stochastically driven nonlinear process in the passively mode-locked laser fiber cavity [25]. Meanwhile, such fluctuations are also shown in the corresponding temporal trains shown in Fig.3 (a).

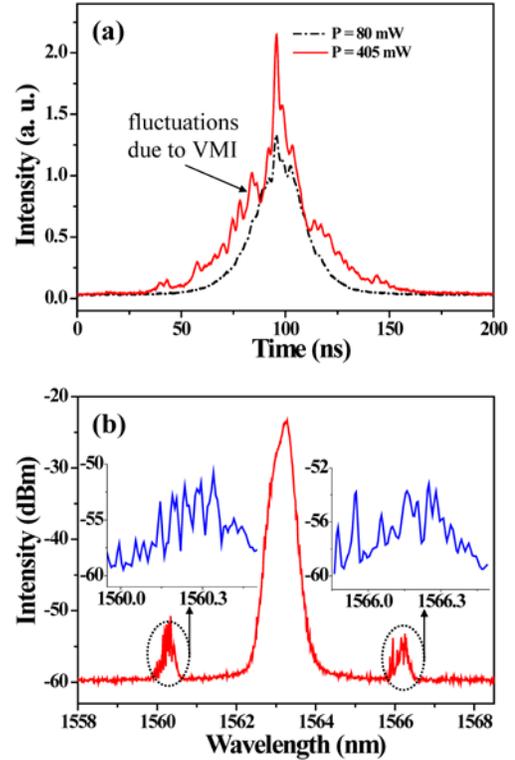

Fig. 3. (a). Temporal output for pump power at 80 mW and 405 mW, respectively and (b) optical spectrum for pump power at 405 mW, and the insets show the fluctuations on the generated sidebands.

## III. DISCUSSIONS

The vector nature of the ML is checked with a polarization beam splitter (PBS) to separate completely the orthogonal polarization components of the pulse into two polarization maintaining fibers. We define one port of the PBS as x polarization, and the other as y polarization. Due to the connection loss between the output fiber and the PBS, noise is shown in one port of the PBS, although it can be neglected.

Figs. 4 (a) and (b) show the two orthogonal components for pump at 150 mW and 405 mW, respectively. As shown, the two components are almost identical for laser without XPMI. However in the presence of XPMI they are different with 3 features: (1). the center wavelength of the two polarization components (Fig. 4 (b)) is ~0.08 nm, while zero for laser without XPMI (Fig. 4 (a)); (2). their corresponding temporal outputs of the two polarization components are slightly different (Fig.4 (c)); (3). sidebands are symmetrical in frequency, but with different intensities (Fig. 4 (b)).

The main reason for features (1) and (2) is the formation of DFWM. As we can see, the two polarization components of the spectrum are almost identical when no DFWM happens, while they diverge gradually when increasing pump power. Through DFWM process, part of energy of the center wavelength is converted into the sidebands. As the phase matching conditions of the two polarization directions are slightly different, that's why the DFWM efficiencies are also different. As a result, both shape and intensity of the polarization components of the center wavelength are different, as well as the temporal outputs.

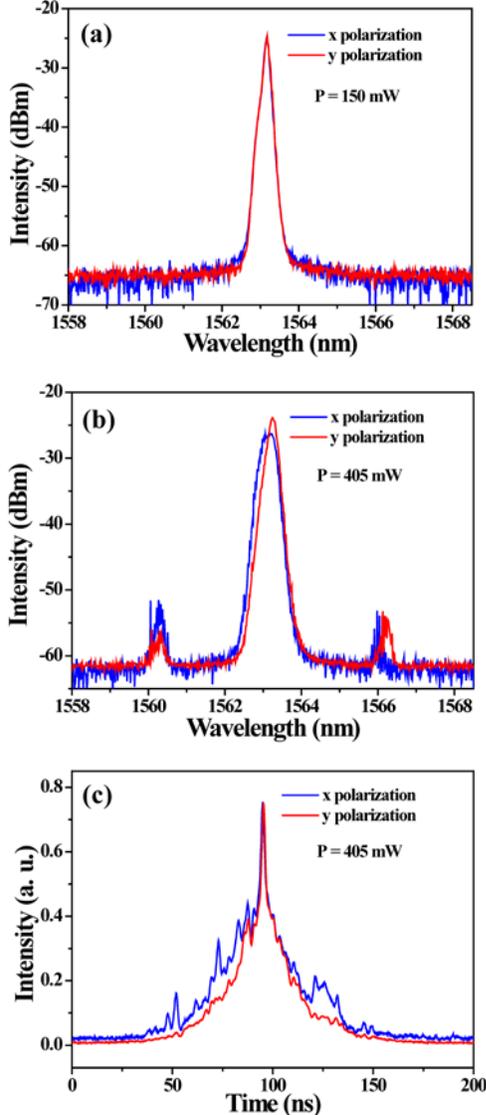

Fig. 4. (a). Two orthogonal components of optical spectra for pump at 150 mW and (b) 405 mW. (c). corresponding temporal trains for pump at 405 mW.

The frequency symmetry of the sidebands is a result of the energy conservation of DFWM, while the intensity asymmetry can be explained by the XPMI of two components in an elliptical polarized laser. The intensity difference between the sidebands in one polarization direction proves that neither a scalar MI nor a polarization MI, but a XPMI contributes to the difference since the intensities of the sidebands are determined by the powers of center wavelength in both two directions. As the laser is not circularly polarized, the intensities of the sidebands generated via DFWM are also different.

The polarization state evolution under different pump strengths is shown in Fig. 5. During the experiment, the measured polarization state may be distorted by the 1 m output SMF. However once the total system is fixed, the output polarization state would follow the true change of the polarization state. As shown in Poincaré sphere, when the pump power is relatively low, the polarization state of the ML is unstable. Further increased pump strength results in ML with more stable polarization state, which evolves from nearly circularly polarization to elliptical polarization. However, DFWM process emerges even in such a severe polarization variation. The possible reason is that the phase matching condition of this spontaneous DFWM in a ML cavity is easier to be satisfied than that pumped directly by a train of pulses in a section of fiber. This is understandable in that the pulses would circulate thousands of times in the ring cavity.

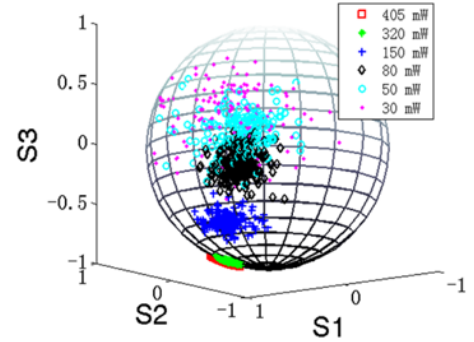

Fig. 5. Polarization states under different pump strengths.

The polarization variation, mainly from rotating PC mechanically, influences the operating state of the laser for two reasons: (1) the saturable absorber characteristics of graphene is polarization-dependent; (2) the birefringence and polarization state variation induced by PC rotation changes the phase matching conditions of nonlinear process in the fiber ring cavity. As we cannot control the polarization state quantitatively due to the mechanically PC rotation method, only a brief description is given here. Depending on different PC settings, we find that the MI process can either be scalar or vector. For laser functioning in XMPI, a small PC rotation changes the intensities of two sidebands. In some cases, laser functions in scalar MI, and two sidebands with equal intensity are much narrower and higher.

Although XPMI has been reported in net anomalous dispersion fiber [26] where both dispersion, gain and loss, nonlinearity, and SA property have influences on the XPMI, in the experiment, we did not observe MI with net anomalous dispersion. The possible reason is that anomalous dispersion broaden a pulse width, leading to a smaller peak power, while normal dispersion compresses a pulse width and its peak power may be large enough to generate new spectral components due to nonlinear process when we increasing pump strength. So, it is possible to observe MI in net anomalous dispersion cavity when other parameters are settled properly.

Here, the pulse train is generated in the ML cavity with fundamental frequency, so the pulse period is equal to the round-trip time of the ring cavity. That's to say, the system is synchronously pumped as in [15]. Different from [27], we do not observe Raman-Stokes wave, therefore no additional sidebands ever emerge except for two orthogonal XPMI sidebands. This is because the birefringence in our experiment is low, and the group dispersion mismatch is much smaller than that in [27].

Limited by the power of the pump laser, a maximum sideband gain of 9 dB is obtained, however, they can grow higher. The key to enhance the efficiency is to increase the nonlinearity, which is mainly determined by the taper-graphene structure. Therefore, we can improve the nonlinearity via increasing the taper length, reducing the taper waist diameter, and adding another high nonlinearal fiber. Other methods include optimizing polarization state of the cavity, reducing the insertion loss, and increasing the pump strength. As the reGO is produced by reducing graphene oxide, defects are found in this material. It is also useful to improve the efficiency of the sidebands by utilizing graphene produced from mechanical exfoliation method or chemical vapor deposition.

As two new photons with orthogonally perpendicular polarization states are generated simultaneously, they are naturally related [11]. Our scheme provides a direct bright polarization-correlated photon source for quantum communication, and no other equipments for generating entangling state are needed. Compared with correlated photons created by spontaneous parametric down conversion [28, 29], our cavity with spontaneous DFWM is rather efficient and communication compatible. Our design can provide new devices in quantum communication, wavelength conversion and parametric amplifiers.

## IV. CONCLUSION

We experimentally observe XPMI in ML due to enhanced nonlinearity based on a long fiber taper deposited with reGO. Results show that two polarizations are excited, and they interact with each other via cross phase modulation. The phase matching of the XPMI is satisfied even the polarization state of the laser changes greatly under different pump strengths. This scheme would find potential applications in quantum communication, wavelength conversion and parametric amplifiers.